\documentclass[twocolumn,showpacs,preprintnumbers,amsmath,amssymb,superscriptaddress]{revtex4-1}

\usepackage{graphicx}
\usepackage{dcolumn}
\usepackage{bm}

\begin{document}


\title{Experimental Evidence of Fluctuation-Dissipation Theorem Violation\\ in a Superspin Glass}

\author{Katsuyoshi Komatsu}
\email{katsuyoshi.komatsu@cea.fr}
 \altaffiliation{}
\author{Denis L'H${\rm \hat{o}}$te}%
\email{denis.lhote@cea.fr}
\affiliation{Service de Physique de l'Etat Condens${\acute{e}}$ (CNRS/URA 2464), DSM/IRAMIS/SPEC, CEA Saclay, F-91191 Gif/Yvette Cedex, France.
}%
\author{Sawako Nakamae}%
\affiliation{Service de Physique de l'Etat Condens${\acute{e}}$ (CNRS/URA 2464), DSM/IRAMIS/SPEC, CEA Saclay, F-91191 Gif/Yvette Cedex, France.
}%

\author{Vincent Mosser}
\affiliation{
ITRON SAS, 76 avenue Pierre Brossolette, F-92240 Malakoff, France.
}%

\author{Marcin Konczykowski}
\affiliation{
Laboratoire des Solides Irradi${\acute{e}}$s, Ecole Polytechnique, F-91128 Palaiseau, France.
}%

\author{Emmanuelle Dubois}
\affiliation{
Laboratoire PECSA, UMR 7195 CNRS, Universit${\rm \acute{e}}$ Pierre et Marie Curie, - 4 place Jussieu, Bo${\hat{\imath}}$te 51, 75252 Paris Cedex 05, France.
}%
\author{Vincent Dupuis}
\affiliation{
Laboratoire PECSA, UMR 7195 CNRS, Universit${\rm \acute{e}}$ Pierre et Marie Curie, - 4 place Jussieu, Bo${\hat{\imath}}$te 51, 75252 Paris Cedex 05, France.
}%
\author{Regine Perzynski}
\affiliation{
Laboratoire PECSA, UMR 7195 CNRS, Universit${\rm \acute{e}}$ Pierre et Marie Curie, - 4 place Jussieu, Bo${\hat{\imath}}$te 51, 75252 Paris Cedex 05, France.
}%

\date{\today}

\begin{abstract}
We present the experimental observation of the fluctuation-dissipation theorem (FDT) violation in an assembly of interacting magnetic nanoparticles in the low temperature superspin glass phase. The magnetic noise is measured with a two-dimension electron gas Hall probe and compared to the out of phase ac susceptibility of the same ferrofluid. For ``intermediate'' aging times of the order of 1~h, the ratio of the effective temperature $T_{\rm eff}$ to the bath temperature $T$ grows from 1 to 6.5 when $T$ is lowered from  $T_g$ to 0.3 $T_g$, regardless of the noise frequency. These values are comparable to those measured in an atomic spin glass as well as those calculated for a Heisenberg spin glass.
\end{abstract}

\maketitle

During the last two decades, the extension of the FDT to the out-of-equilibrium regime has been the subject of many theoretical and experimental investigations \cite{Bouchaud92,Parisi97,Cugliandolo97,Marinari97,Barrat98,Sciortino01,Kawamura03,Grigera99,Bellon01,Herisson02,Herisson04,Abou04,Buisson05,Lucchesi06,Greinert06,Strachan06,J-Farouji07,Joubaud09,Jop09,Maggi10,Oukris10}. In the ``weak ergodicity breaking'' scenario \cite{Bouchaud92,Cugliandolo97}, it has been shown that the concept of an effective temperature ($T_{\rm eff}$) \cite{Cugliandolo97} that differs from the bath temperature ($T$) enables the extension of the FDT to the out of equilibrium regime. The FDT violation has been investigated in several numerical simulations \cite{Andersson92,Bouchaud92,Franz95,Parisi97,Barrat98,Sciortino01,Kawamura03}, while experimental studies are rather scarce: They concern one molecular glass \cite{Grigera99}, colloids \cite{Bellon01,Abou04,Greinert06,Strachan06,J-Farouji07,Maggi10}, polymers \cite{Buisson05,Lucchesi06,Oukris10}, one liquid crystal \cite{Joubaud09} and one spin glass (SG) \cite{Herisson02,Herisson04}. On the other hand,the absence of FDT violation is reported in colloids \cite{J-Farouji07,Jop09} and in a magnetic nanoparticle system \cite{Jonsson95,Jonsson98}. Thus, the systems and the conditions in which the FDT is violated still represent an open question.

Here, we investigate the FDT violation in an out-of-equilibrium superspin glass (SSG) system. The magnetic nanoparticles suspended in fluid (glycerol) have a single-domain magnetic structure. Therefore, their magnetic moment of $\sim10^{4}{\mu}_{\rm B}$ behaves as one large spin, and is called a ``superspin''. Once the carrier matrix is frozen, the positions as well as the anisotropy axis orientations of the particles are fixed, and the only remaining degree of freedom is the superspin rotation. The randomness and disorder found in the nanoparticle positions, orientations and sizes lead to magnetically glassy behaviors at low temperatures, including slow dynamics and aging effect, similar to those of atomic SGs, hence these systems are called ``superspin glasses'' \cite{Jonsson95,Mamiya99,Parker08,Wandersman08,Nakamae09,Sun03}. Due to the large magnetic moment, slow correlation length growth, etc., the observation of magnetic noise within experimental frequency/time range becomes more feasible in a SSG system. Furthermore, the much slower microscopic time scale in SSG than that in SG can help to fill the large time scale gap between the computer simulations and experiments.

The FDT describes the relation between the power spectrum of fluctuations of an observable,
$\delta M(\omega)$ (here $M$ is the magnetization) and the imaginary component of the ac susceptibility $\chi ''(\omega)$ to the conjugate field  \cite{Alba87} :
\begin{eqnarray}
\langle\delta M (\omega)^2\rangle=\frac{2k_{\rm B}T}{\pi V}\left(\frac{\chi '' (\omega)}{\mu_0\omega}\right)~\rm{(SI~units)}.
\label{eq:FDT}
\end{eqnarray}
Here, $\langle\cdots\rangle$ denotes the ensemble average per frequency unit, $k_{\rm B}$ is the Boltzmann constant, $T$ the temperature, and $\omega=2\pi f$($f$ is the measurement frequency). The departure from equilibrium can be estimated through the fluctuation-dissipation ratio $X(\omega,t_{\rm w}) = 2k_{\rm B}T\chi''/(\mu_0\omega\langle(\delta M)^2\rangle\pi V)$, or the effective temperature $T_{\rm eff} = T/X(\omega,t_{\rm w})$. $X$ (and $T_{\rm eff}$) depend on $t_{\rm w}$, the waiting time (or the ``age'') at $T$ after a temperature quench from above the glass transition temperature of the system. At equilibrium, the FDT gives $X = 1$ and thus $T_{\rm eff} = T$ while in the aging regime, $X < 1$ and equivalently, $T_{\rm eff} > T$. The effective temperature provides a generalized form of FDT in out-of-equilibrium cases as:
\begin{eqnarray}
\langle\delta M (\omega,t_{\rm w})^2\rangle=\frac{2k_{\rm B}T_{\rm eff}}{\pi V}\left(\frac{\chi '' (\omega,t_{\rm w})}{\mu_0\omega}\right),
\label{eq:FDTaging}
\end{eqnarray}
where $T_{\rm eff}$ rather than $T$ acts as the system temperature, e.g., ``weak ergodicity breaking'' system. Note that in the $1/\omega\ll t_{\rm w}$ limit, the quasi-equilibrium regime is reached \cite{Cugliandolo97}; that is, the FDT relation is recovered and $X=1$.
  
\begin{figure}[h]
\includegraphics[width=3.2in]{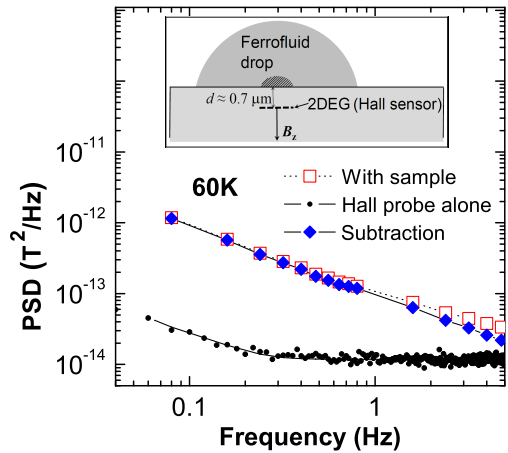}
\caption{\label{fig:noise} Noise power spectrum $S(f)$ of the magnetic field due to the frozen ferrofluid (filled diamonds), obtained by subtracting the Hall probe only spectrum (dots) from the total power spectral density (PSD) (open squares) as a function of frequency  $f$, at 60~K in zero applied field. The power spectral density of the magnetic noise due to the sample was larger than that of the bare Hall sensor by factors of about 25 and 2 at 0.1 and 4~Hz, respectively. Inset: Schematic picture of the magnetic noise measurement setup. The magnetic noise measured in the probe comes mainly from that part of the drop located in front of the 2DEG \cite{LHote09}, indicated by the dark shaded region. (see text).}
\end{figure}

  In this letter, we report the experimental observation of the FDT violation in a frozen ferrofluid in the SSG state via magnetic noise measurements coupled with ac-susceptibility measurements. The ferrofluid used in this experiment is made of maghemite $\gamma{\rm Fe_2O_3}$ nanoparticles dispersed in glycerol with a volume fraction of $15~\%$ in which SSG state has been observed previously  \cite{Parker05,Wandersman08,Nakamae09,Wandersman09}. The particles' average diameter is 8.6 nm and their uni-axial anisotropy energy is $\sim10^{-20}$~J, obtained from the superparamagnetic relaxation time of a diluted sample, $\tau= \bar{\tau}_0\exp(E_{\rm a} /k_{\rm B}T)$ with $\bar{\tau}_0=10^{-9}$~s  \cite{Parker08}, compatible with direct anisotropy field measurements  \cite{Gazeau98}. To measure the magnetic noise, a small drop of ferrofluid was deposited directly onto a Hall probe \cite{LHote09,Komatsu10} (see inset in Fig.~\ref{fig:noise}). All measurements were made well below 190~K, the freezing temperature of glycerol. In a frozen sample, the magnetic moments (superspins) interact with one another through dipolar interactions leading to a static superspin-glass transition at $T_g\sim67$~K \cite{Wandersman08}. The ac susceptibility of the bulk ferrofluid sample (approximately 5 $\rm{\mu l}$) was measured with a commercial SQUID magnetometer. The magnetic noise was measured with a two-dimension electron gas (2DEG) quantum well Hall sensor (QWHS) based on pseudomorphic AlGaAs/InGaAs/GaAs heterostructure with a high mobility and a large Hall coefficient $R_{\rm H}$ ($\sim800$~$\rm{\Omega}$/T). The QWHS has a nominal sensitive area of $\sim2\times2$~${\rm \mu m^2}$, located at $d\sim0.7$~${\rm \mu m}$ beneath the probe surface (see inset in Fig.~\ref{fig:noise}). The ferrofluid drop of about 7~pl has a diameter $\sim30$~${\rm \mu m}$, much larger than the probe sensitive area. We have made use of the spinning current technique which effectively suppresses both the offset and the low frequency background noise of the Hall probe simultaneously \cite{Kerlain08}. In this method, the directions of the current injection and the Hall voltage detection in Hall cross are continuously switched at a spinning frequency, $f_{\rm spin}$ which is larger than the largest noise frequency of interest. Low frequency background noise ($f < 10$~Hz) suppression is of great importance because the typical time scales involved in the fluctuation dynamics of a SSG system are much larger than 1~s. With $f_{\rm spin} = 1 $~kHz, we achieved a field sensitivity of $\sim2 {\rm mG/\sqrt{Hz}}$ (for $f \sim0.1$~Hz) for the temperature range between 20 and 85~K; a 10-fold improvement with respect to the sensor sensitivity obtained without this technique. The noise power spectra $S(f)$ of the magnetic field were measured in two distinct frequency regions; from 0.08 to 0.7~Hz and from 0.8 to 8~Hz. All magnetic noise data of the ferrofluid (except at 85~K) were 
taken following a temperature quench from 85~K (= 1.27~$T_g$) to the measurement temperatures and a waiting time of 10 minutes for temperature stabilization. Figure \ref{fig:noise} shows an example of such a spectrum, taken at 60~K. $S(f)$ is calculated via $S(f) = \langle[\delta B_{\rm z}( f )]^2\rangle = ( I R_{\rm H} )^{-2}  \langle(\delta V_{\rm H})^2\rangle$, where $\delta V_{\rm H}$ is the fluctuation of the measured Hall voltage, $\delta B_{\rm z}$ is the corresponding fluctuation of the (uniform) field $B_{\rm z}$ perpendicular to 
the Hall probe and $I$ the injection current. Here the symbol $\langle\cdots\rangle$ indicates an averaging over a large data set. Each spectrum was obtained from averaging over 300 and 3000 spectra in the low and high frequency regions, respectively. The aging time $t_{\rm w}$ of the system is thus this averaging time, which is always of the order of a few $10^3$~s. This is an ``intermediate'' waiting time used in typical aging experiments on bulk ferrofluid SSG samples where $t_{\rm w}$'s range from a few $10^2$~s to several $10^4$~s  \cite{Wandersman08}.

\begin{figure}[h]
\includegraphics[width=3.2in]{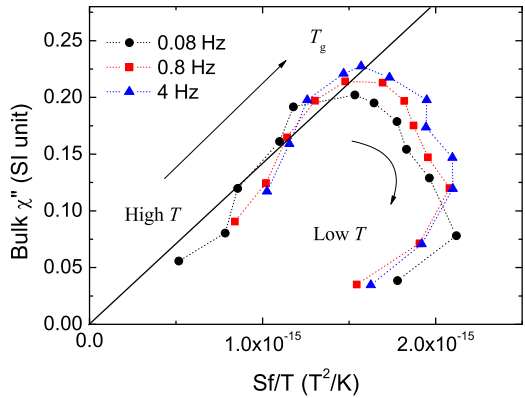}
\caption{\label{fig:FDTviolation} $\chi ''(f, T)$ of bulk sample as a function of $S(f,T)f/T$ for frequencies, 0.08, 0.8 and 4~Hz. Each data point corresponds to $\chi ''$ and $S$ measurements at a given bath temperature $T$ and frequency $f$. The solid straight line indicates the linear relation in the high temperature region above $T_g$.}
\end{figure}

\begin{figure}[h]
\includegraphics[width=3.2in]{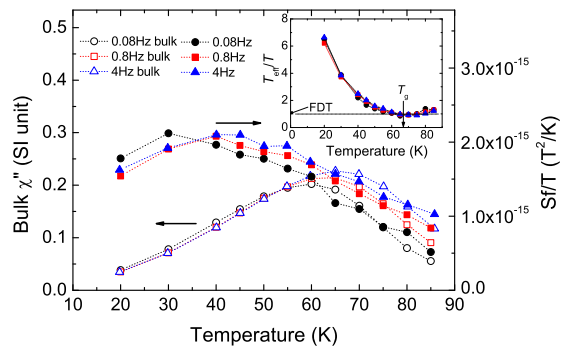}
\caption{\label{fig:Tdependence} Temperature dependent $\chi ''$ of bulk sample (open symbols) and $Sf/T$ (filled symbols) at frequencies; 0.08, 0.8 and 4~Hz. The relative normalization of the two vertical scales corresponding to $\chi ''$ and $Sf/T$ is given by the slope of the straight line in Figure \ref{fig:FDTviolation}. Inset: The temperature dependence of $T_{\rm eff}/T$ at $f = 0.08$, 0.8 and 4~Hz. The horizontal line corresponds to the FDT relation, i.e., $T_{\rm eff} / T = 1$.}
\end{figure}

Figure \ref{fig:FDTviolation} shows the imaginary part of the ac magnetic susceptibility $\chi '' (f, T)$ of a bulk sample as a function of $Sf/T$ at $f= 0.08$, 0.8, 4~Hz. $\chi ''(f, T)$ at each temperature was measured with the aging time $t_{\rm w}$ of 1 hour after the temperature quench from 85~K. We found that all data points collected above $T_g = 67$~K are aligned along a common straight line; i.e., $\chi '' \propto Sf/T$. The solid straight line in Fig.~\ref{fig:FDTviolation} is the best fit to these data points for $T = T_g$ for all three frequencies. This linear relationship is independent of $f$, indicating that the FDT holds between the two quantities in this $T$ range according to Eq.~\ref{eq:FDT}. The data points deviate from the straight line starting from the maximum value of $\chi ''$ occurring near $T = T_g$ and downwards in temperature. Figure \ref{fig:Tdependence} shows the temperature dependencies of $\chi ''$ and $Sf/T$ (same data as in Fig.~\ref{fig:FDTviolation}). The relative normalization of the two vertical scales, $\chi ''$ and $Sf/T$, is given by the slope of the straight line found in Fig.~\ref{fig:FDTviolation}. As can be seen from the figure, $\chi ''$ and $Sf/T$ superpose in the high temperature region above $T_g$, while they separate below $T_g$. The deviation from the linear relation and the separation of the normalized $\chi ''$ and $Sf=T$ below $T_g$ indicate a clear departure from FDT.
The slope value, $\chi '' /(Sf/T)=(1.4\pm 0.2)\times10^{14}$~[K/$\rm T^2$] in the high temperature region (see Fig.~\ref{fig:FDTviolation}) is determined by the effective volume $V_{\rm eff}$ of ferrofluid that contributes to the magnetic noise measurement \cite{LHote09} and by the magnitude of the magnetic field induced by the ferrofluid in the Hall probe. Because of the sample geometry and of the $1/r^4$ \cite{LHote09} dependence of the dipolar field variance $\langle(\bar{\delta B_{\rm z}})^2\rangle$, where $\bar{B_{\rm z}}$ is the  average of $B_{\rm z}$ induced by the sample over the probe sensitive area, $V_{\rm eff}$ is confined within a volume close to the sensor surface (see inset of Fig.~\ref{fig:noise}). 
To check the quantitative consistency of the above analysis, we have estimated the slope value independently. In depth investigations of the response of a Hall cross to an inhomogeneous perpendicular field $B_{\rm z}$ have revealed that this response is proportional to the average of $B_{\rm z}$ over the effective area $a_{\rm eff}$ of the probe which is about twice the junction area, i.e., $a_{\rm eff}=2w^2$ ($w$ being the width of the cross arms) \cite{Bending97,*Ibrahim98}. We evaluated numerically the variance $\langle(\bar{\delta B_{\rm z}})^2\rangle$ with $B_{\rm z}$ being the sum of contributions from elementary volumes $d^3\mbox{\boldmath$r$}$ of the sample, each having a magnetic moment variance given by FDT; that is ($2k_{\rm B}T\chi ''/\pi \mu_{0}\omega)d^3\mbox{\boldmath$r$}$. 
The calculated slope is $(0.7\pm0.25)\times10^{14}$~[K/$\rm T^2$]. The uncertainty comes mainly from that of the response function of the probe, which is partly due to the uncertainty in the true value of $w$ (1 $\mu {\rm m} < w < 2 \mu {\rm m}$) caused by the edge depletion effect. Another source of uncertainty comes from the fact that the effect of averaging $B_{\rm z}$ over the probe area has been evaluated using a Monte-Carlo simulation to which some simplifying assumptions were made, i.e., independent superspins, square probe area, etc. Despite these elements taken into account, the measured and calculated slope values are close to each other, lending credibility to our results.

Below the SSG transition temperature $T_g$, where the system is in an out-of-equilibrium state, we have witnessed a departure from the equilibrium FDT relation. We now estimate the effective temperature $T_{\rm eff}$ as evoked above from the FDR ratio of $\chi ''$ to $Sf/T$ (see Eq.~\ref{eq:FDTaging}). The inset in Fig.~\ref{fig:Tdependence} shows the temperature dependence of $T_{\rm eff} / T$ obtained at $0.08, 0.8$ and $4$~Hz. $T_{\rm eff}/T$ increases monotonically when $T$ decreases, starting from 1 around $T_g$, to 6.5 at $0.3T_g$ (= 20~K) regardless of the frequency. The values of $T_{\rm eff}/T$ are of the same order as those reported in the experimental study of an atomic SG, $T_{\rm eff}/T = 2.8 - 5.3$ \cite{Herisson04} and in a Monte-Carlo simulation on a Heisenberg SG, $T_{\rm eff}/T = 2 - 10$ \cite{Kawamura03}.

The observation of $T_{\rm eff} > T$ suggests that the system is in the aging regime, i.e., not in the so-called quasi-equilibrium regime \cite{Cugliandolo97} where observation times $t_{\rm obs}=2\pi/\omega$ are much smaller than the aging time $t_{\rm w}$. Here, $t_{\rm obs}\sim 1$~s is rather short compared to $t_{\rm w}\sim10^3$~s, corresponding to $t_{\rm obs}/t_{\rm w}\sim 10^{-3}$. 
Violations of FDT have been observed experimentally for very low values of $t_{\rm obs}/t_{\rm w}$: $10^{-7}$-$10^{-4}$ in a molecular glass \cite{Grigera99}, $10^{-5}$-$10^{-3}$ in polymer glasses \cite{Buisson05,Lucchesi06}, and $10^{-7}$-$10^{-4}$ in colloidal glasses \cite{Bellon01,Abou04,Maggi10}.  Furthermore in those experimental systems, $T_{\rm eff}$ does not rapidly approach the bath temperature $T$ with waiting time $t_{\rm w}$. Through numerical simulations on domain growth systems \cite{Barrat98}, the breaching of the quasi-equilibrium state depends on the system itself and on the two time scales ($t_{\rm obs}$ and $t_{\rm w}$) separately rather than on $t_{\rm obs}/t_{\rm w}$  \cite{Cugliandolo02}. Similar conclusions were drawn in SG simulations \cite{Andersson92,Franz95}.
In an interacting magnetic nanoparticle SSG system similar to ours, the FDT remained valid for $t_{\rm obs}/t_{\rm w}<10^{-5}$ \cite{Jonsson98}. Thus, it is tempting to conjecture that the limit between the two regimes lie somewhere between $t_{\rm obs}/t_{\rm w}=10^{-5}$ and $10^{-3}$. However, one must be careful because the differences between the two systems (particle sizes, concentrations, etc.) and their experimental conditions (measurement techniques, temperature quench protocol, etc.) do not allow direct comparison between the two studies. Comparing the SSG and SG systems, we note that the interaction between superspins is of the long range dipolar type whereas between atomic spins, it is of the short range exchange type \cite{Franz95,Herisson02,Herisson04}. 
Thus far, a large scale dynamical simulation on nanoparticle systems with random anisotropy has not been investigated in terms of the FDT relation. Comparisons of experimental data to such simulation result will be very interesting.

  In conclusion, we have presented an experimental evidence of FDT violation in the out-of-equilibrium, aging SSG state of a frozen ferrofluid through magnetic noise measurements. For an aging time of about 1 hour, the extracted effective temperature (normalized to the bath temperature), increases by a factor of 6.5 when $T$ decreases from $T_g$ to $0.3T_g$. Such values are of the order of those found in an atomic SG \cite{Herisson04} and in a numerical simulation of a Heisenberg SG \cite{Kawamura03}. More investigations are needed to elucidate aging time dependence of $T_{\rm eff}$. 

We thank R. Tourbot for precious technical help and L. Cugliandolo, J. Kurchan, F. Ladieu, S. Franz, A. Barrat and E. Vincent for illuminating discussions. This work was supported by Triangle de la Physique (contracts MicroHall and DynMag). 


%

\end{document}